\documentclass{article} 
\usepackage{iclr2026_conference,times}

\usepackage{amsmath,amsfonts,bm}

\def\eqref#1{equation~\ref{#1}}

\def\1{\bm{1}}

\DeclareMathAlphabet{\mathsfit}{\encodingdefault}{\sfdefault}{m}{sl}
\SetMathAlphabet{\mathsfit}{bold}{\encodingdefault}{\sfdefault}{bx}{n}

\usepackage{amssymb}
\usepackage{amsmath}
\usepackage{hyperref}
\usepackage{url}
\usepackage{microtype}
\usepackage{lettrine}
\usepackage{xltabular}
\usepackage{booktabs}
\usepackage{xspace}
\usepackage{colortbl}
\usepackage{multirow}
\usepackage{cleveref}
\usepackage{pdfpages}
\usepackage{algorithm}
\usepackage{algpseudocode}
\usepackage{stmaryrd}
\usepackage{tabularx}
\usepackage{subcaption}

\usepackage{pgfplots}
\usepgfplotslibrary{colormaps}
\pgfplotsset{
    compat=newest,
	colormap={justblackandwhite}{color=(white) color=(white) color=(white)}
}

\definecolor{coverred}{RGB}{29, 49, 43} 
\definecolor{customgreen}{RGB}{193,199,205}
\definecolor{customblue}{RGB}{166,200,254}
\definecolor{mygreen}{RGB}{204, 255, 224}
\definecolor{myorange}{RGB}{252, 225, 187}
\definecolor{myred}{RGB}{247, 185, 185}
\definecolor{iccvblue}{rgb}{0.21,0.49,0.74}
\usepackage[detect-all,locale=UK,per-mode=symbol]{siunitx}
\DeclareSIUnit\pixel{px}

\usepackage{standalone}
\usepackage{tikz}
\usepackage[toc,page]{appendix}

\usetikzlibrary{fadings}
\newcommand\insarflow{\texttt{InSARFlow}}

\newcommand{\ie}{\textit{i.e.,}}

\title{InSARFlow: Riemannian Flow Matching for Interferometric SAR}

\author{Georges Le Bellier\\
Inria, ENS de Lyon, CNRS, Université Claude Bernard Lyon 1, LIP, Lyon, France\\ 
CEDRIC, CNAM, Paris, France \\
\AND
Arnaud Breloy \\
CEDRIC, CNAM, Paris, France \\
\AND
Nicolas Audebert \\
 Université Gustave Eiffel, Géodata Paris, IGN, LASTIG, Marne-la-Vallée, France\\
CEDRIC, CNAM, Paris, France \\
}

\iclrfinalcopy 
\begin{document}

\maketitle

\fancyhead{}

\begin{abstract}
%
Interferometric synthetic aperture radar (InSAR) enables the analysis of topological changes in natural sites or urban areas.
This technique operates by recovering the surface displacement from the study of the phase difference (interferogram) between two or multiple SAR images of the same scene at different times.
In practice, the use of phase denoising algorithms is necessary to extract relevant information from the interferograms, which motivated numerous developments in the past decade. 
Due to the inherent periodicity of phases, interferograms can naturally be represented as points on a Riemannian manifold (specifically, the flat torus).
This strong prior knowledge motivates the design of learning-based methods that respect the interferogram's intrinsic geometric constraints.
Building on this observation, this paper introduces \texttt{InSARFlow}: a generative modeling framework for the denoising and generation of interferograms.
Unlike conventional deep learning approaches formulated in Euclidean spaces, which may violate phase periodicity and produce invalid estimates, this method leverages \textit{Riemannian flow matching} to learn distributional transports directly on the torus. 
The interest of this approach is demonstrated on two use cases: 
($1$) realistic interferogram generation, in which we model the transport between uniform noise on the torus to real-world interferograms;
($2$) interferogram denoising, where \insarflow{} transports noisy interferograms to their denoised counterparts.
Extensive experiments on a controlled synthetic dataset validate the effectiveness of \insarflow{} for interferogram denoising.
The proposed approach consistently outperforms studied baselines, highlighting the benefits of geometry-aware learning on the torus.

\noindent Code and weights available at \url{https://github.com/lebellig/insarflow}.

\end{abstract}

\section{Introduction}



InSAR was employed for surface deformation measurements and elevation estimations starting from the 90's \cite{Gabriel1989, Massonnet1993}, and remains an effective geodetic tool today, due to its powerful and unique capability \cite{InsartheoryESA, InSARerroranalysis_Hanssen, Ferrettibook}.
In particular, remote sensing missions (Sentinel-1, UAVSAR, TerraSAR-X, etc.) have provided unprecedented coverage of Earth imagery using SAR (sometimes up to the sub-centimeter level of accuracy).
These acquisitions, taken at regular intervals, make it possible to accurately monitor the deformation of Earth’s surface.
Applications include, for example, tracking the subsidence of urban areas due to construction work~\cite{groundInsar}, or analyzing subtle changes in the topography of natural sites such as volcanoes~\cite{rs12061019, DITRAGLIA2021112441}.
In this context, deformation information lies in the phase difference between the (complex-valued) pixels of two georeferenced and co-registered SAR images.
An interferogram represents the phase difference between two co-registered SAR images acquired on different dates and contains information about the geometry of the observed area.
In practice, the loss of coherence between the signals at two dates generally induces a high level of noise, and interferograms have to be denoised before any displacement estimation. 
Since SAR images exhibit particular noise statistics, this InSAR phase denoising step motivated numerous developments \cite{xu2020insar}.
Two major families of approaches can be identified.
The first one treats single images of phase differences and aim at spatially smoothing noisy observations of interferometric fringes.
These methods range from traditional image processing \cite{song2017insar, deledalle2010nl} to more recent deep learning based solutions \cite{insarmonet_vitale_2022, insar2insar_geara_2024}.
The second one is multitemporal, taking advantage of all images that may have been acquired between the two dates in question to produce a more accurate estimate of the phase shift \cite{guarnieri_2008, ferretti_2011, fornaro2014caesar, ning_cao_2015, Ansari2018}.
A prominent method in this family is referred to as \textit{interferometric phase linking}, and aims at recovering consistency between all possible interferograms across time (cf. \cite{minh2023interferometric, vu2025covariance} for recent overviews).

In this context, several attempts have explored the use of generative models for InSAR through adaptations of adversarial methods, primarily addressing the phase unwrapping problem~\cite{pu-gan, unwrap_pix2pix} and interferogram generation~\cite{zhou2024ganinsar}. 
However, recent advances in generative modeling based on diffusion models~\cite{ho2020denoising} and flow-based models~\cite{albergo2023stochastic, lipman2022flow} have demonstrated remarkable generative capabilities across a wide range of modalities, including images, molecular structures, and text. 
Moreover, these models exhibit strong connections with denoising objectives \cite{vincent, gagneux2025generation}.
Their application to InSAR therefore represents a highly promising yet still underexplored research direction. 
To the best of our knowledge, diffusion models have been applied so far to InSAR mainly for the generation of unwrapped InSAR interferograms~\cite{TuelKerdreux2023}, or for phase unwrapping~\cite{song2025unwrapdiff}.
The main bottleneck to the application of flow-based models to InSAR lies in the nature of the image space, as SAR image pixels are complex-valued and, in InSAR, the relevant information is encoded in their phase. 
Since deep learning models usually operate in Euclidean spaces, they may produce outputs that do not lie on the phase manifold, \ie values outside the interval $[0, 2\pi)$. 
Moreover, they can be inefficient, as part of their capacity is spent learning the structure of the data manifold itself. 
This motivates the use of deep models operating directly on the phase manifold.

\begin{figure}[t]
    \centering
    \input{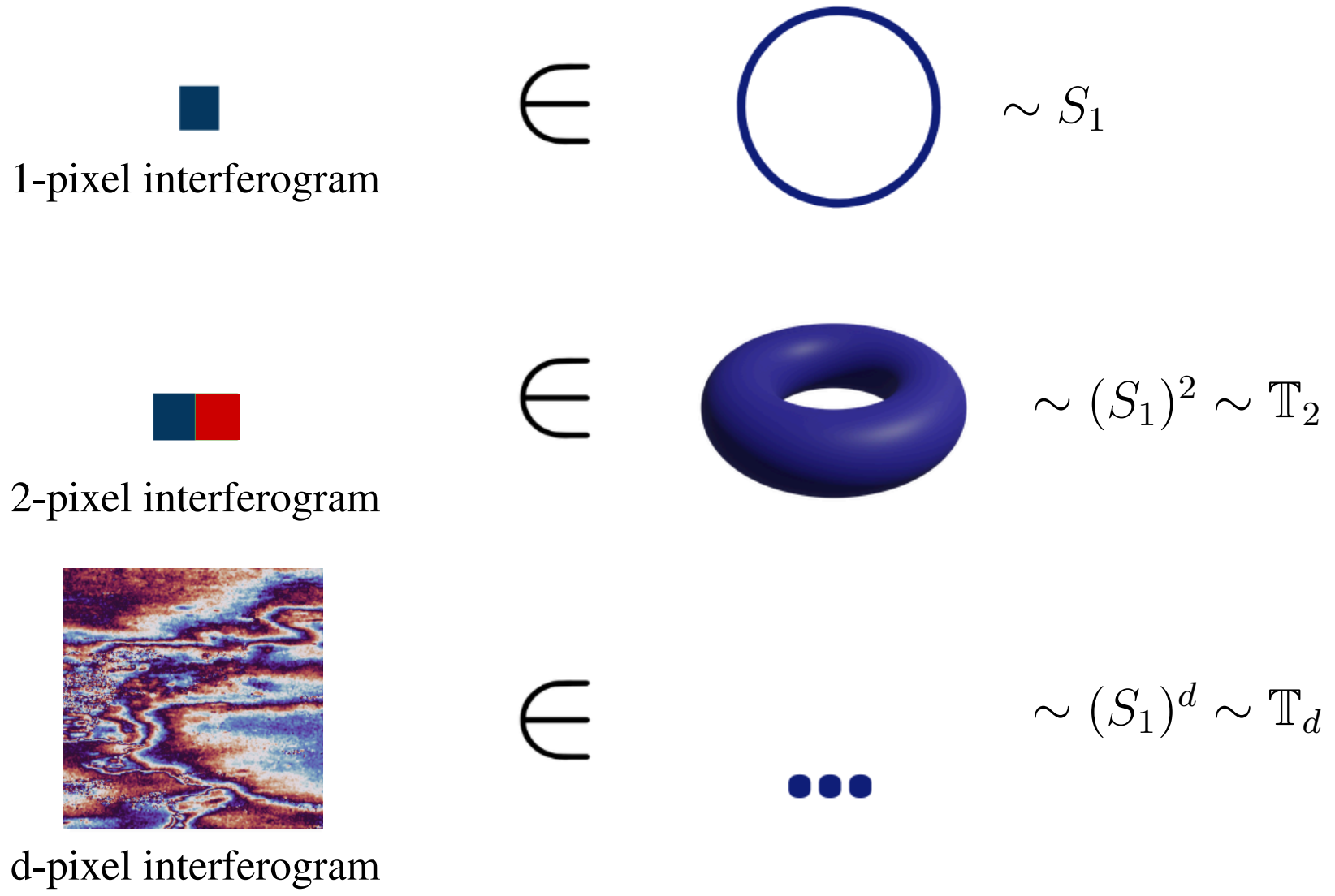}
    \caption[Periodicity of interferograms.]{Periodicity of interferograms. A 1-pixel interferogram lies on the unit circle $S_1$; a 2-pixel interferogram lies on the 2-torus $\mathbb{T}_2 = (S_1)^2$; and, more generally, a $d$-pixel interferogram lies on the $d$-dimensional torus $\mathbb{T}_d = (S_1)^d$.}
    \label{fig:datamanifold}
\end{figure}

In this work, we address the aforementioned issues from a geometric perspective. 
By interpreting a patch of an InSAR phase image as a point on a high-dimensional torus (\cref{fig:datamanifold}), we can leverage \textit{Riemannian} flow models~\cite{chen2024flow}. 
The generative process then operates directly on the phase manifold, ensuring the validity of the generated interferograms. In contrast to deep models operating in Euclidean space, where no such guarantee exists, it is ensured here by construction.

\begin{figure}[!t]
    \centering
    \includegraphics[width=\linewidth]{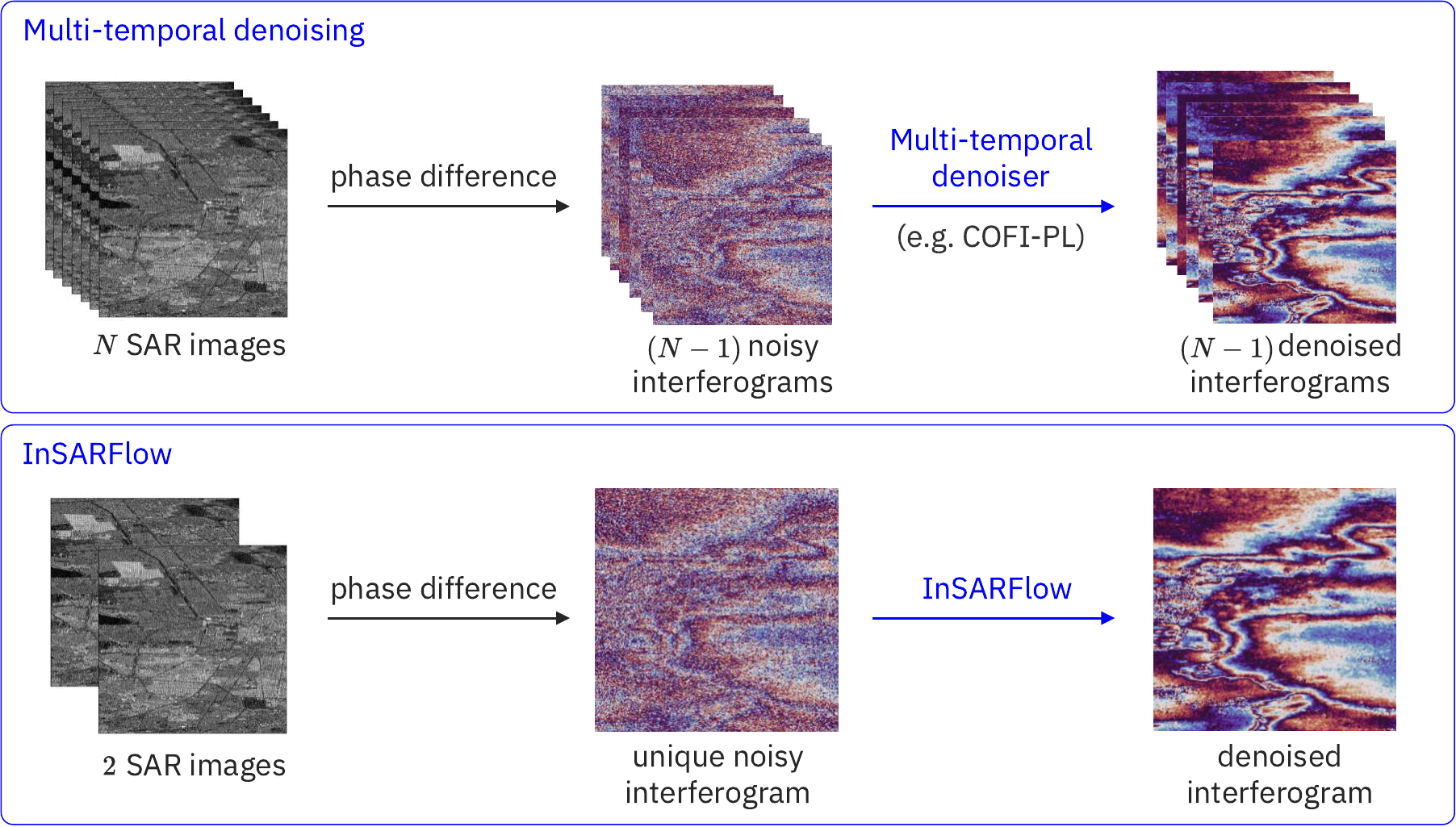}
    \caption[InSARFlow, a Riemannian flow matching model, learns to predict, from a single noisy interferogram $\mathbf{w}$, its denoised version.]{InSARFlow, a Riemannian flow matching model, learns to predict, from a single noisy interferogram $\mathbf{w}$, its denoised version, estimated by a multitemporal denoiser. In contrast to the multitemporal denoiser, which requires the full timeseries of interferograms, InSARFlow directly processes single interferograms.}
    \label{fig:framework}
\end{figure}

Armed with these tools, we then consider two problems:
\begin{enumerate}
    \item \textbf{The generation of interferograms for realistic simulation purposes.}
    We transport a prior distribution on the torus to the distribution of real data.
    In this setting, we put a particular focus on showing that Euclidean-based method are prone to generate invalid samples, while \insarflow{} only generates valid and realistic interferograms.


\item \textbf{The denoising of InSAR interferograms.} Unlike most state-of-the-art approaches, which are trained and evaluated on simulated datasets with controlled ground truth and noise levels~\cite{insar2insar_geara_2024}, we show the effectiveness of \insarflow{} on \textit{real} interferograms.
Since no ground truth is generally available in this setting, we use interferograms denoised by a multitemporal method applied to the full timeseries as surrogate ground truth (cf. \cref{fig:framework}) and train the model in a distillation framework.
In this way, we obtain a single-interferogram (two-pass InSAR) model that can efficiently denoise real InSAR images, even when a full timeseries is not available.
In addition, we evaluate \insarflow{} against competing baselines on a synthetic dataset for which clean ground-truth interferograms are available to enable a rigorous quantitative assessment.

\end{enumerate}
\newpage


\section{Background}

\subsection{InSAR denoising}

An interferogram, denoted $\mathbf{w}\in [0, 2\pi)^{h\times w}$, represents the phase difference between two co-registered SAR images (of height $h$ and width $w$) acquired on different dates and contains information about the geometry of the observed area. 
If the area is unstable, interference fringe patterns typically appear, which, once unwrapped, allow for the estimation of surface deformation. 
Other physical phenomena affect this measure, resulting in the formula~\cite{rosenInSAR}:

\begin{equation}
\mathbf{w} = \Delta \Phi = 
\underset{\substack{\text{\color{green!60!black}surface} \\ \text{\color{green!60!black}displacement}}}{ \colorbox{mygreen}{$\Delta \Phi_\text{disp}$} }
+ 
\underset{\substack{\text{\color{orange!80!black}removed with} \\ \text{\color{orange!80!black}preprocessing}}}{ \colorbox{myorange}{$\Delta \Phi_\text{topo} + \Delta \Phi_\text{orb} + \Delta \Phi_\text{atm}$} }
+ 
\underset{\substack{\text{\color{red!80!black}corruption} \\ \text{\color{red!80!black}noise}}}{ \colorbox{myred}{$\Delta \Phi_\text{noise}$} } [2\pi]
~.
\end{equation}
The terms $\Delta\Phi_\text{topo}$, $\Delta\Phi_\text{atm}$, and $\Delta\Phi_\text{orb}$ denote, respectively, the topographic phase component related to terrain elevation, phase variations induced by atmospheric conditions, and phase contributions arising from differences in satellite orbits between the two acquisitions. These orbital and topographic components can be mitigated by careful data preprocessing~\cite{rosenInSAR, elhajjar2025thesis}. However, removing the atmospheric component remains challenging; we therefore assume that its effect is negligible or can be incorporated as part of the corruption noise.
The remaining terms are $\Delta\Phi_\text{disp}$, which encodes the surface displacement of interest, and a noise component $\Delta\Phi_\text{noise}$, which we aim to suppress. By a slight abuse of terminology, we subsequently refer to these preprocessed phase images as \textit{interferograms}, such that only the displacement and noise contributions remain, i.e., $\mathbf{w} = \Delta\Phi_\text{disp} + \Delta\Phi_\text{noise}$.

\begin{figure}[!t]
    \centering
    \begin{minipage}[c]{0.52\textwidth}
        \centering
        \captionof{table}{Notations and conventions used.}
        \label{tab:object_space_riem}
        \setlength{\tabcolsep}{4pt}
        \small
        \resizebox{\linewidth}{!}{%
        \begin{tabular}{lcc}
        \toprule
        & Euclidean & Riemannian \\
        \rowcolor{customblue!50} Name & \multicolumn{2}{c}{Data point}\\
         Notation & $\textbf{x}_t$ & $\textbf{w}_t$\\
        Space & $\mathbb{R}^d$ & $\mathbb{T}_d$ \\
        \rowcolor{customblue!50} Name & \multicolumn{2}{c}{Velocity}\\
        Notation & $v_t(\textbf{x}_t)$ & $v_t(\textbf{w}_t)$\\
        Space  & $\mathbb{R}^d\rightarrow \mathbb{R}^d$& $\mathbb{T}_d\rightarrow T_{\textbf{w}_t}\mathbb{T}_d$\\
        \rowcolor{customblue!50} Name & \multicolumn{2}{c}{Interpolant}\\
        Notation & $\textbf{x}_t = (1-t)\textbf{x}_0 + t\textbf{x}_1$ & $\mathbf{w}_t
            = \exp_{\mathbf{w}_1}\big(\kappa(t) \log_{\mathbf{w}_1}(\mathbf{w}_0)\big)$\\
        Space & $\mathbb{R}^d$ & $\mathbb{T}_d$ \\
        \midrule
        \end{tabular}%
        }
    \end{minipage}\hfill
    \begin{minipage}[c]{0.46\textwidth}
        \centering
        \begin{tikzpicture}
        \begin{axis}[
            axis equal image,
            hide axis,
            z buffer = sort,
            view = {0}{20},
            scale = 1.0
            ]
            \addplot3[
                surf,
                shader    = faceted interp,
                samples   = 20,
                samples y = 40,
                domain    = 0:2*pi,
                domain y  = 0:2*pi,
                colormap name = justblackandwhite, thin
            ] (
                {(3.5+sin(deg(\x)))*cos(deg(\y))},
                {(3.5+sin(deg(\x)))*sin(deg(\y))},
                {cos(deg(\x))}
            );
        	\addplot3[mark=*,mark size=0.5pt, only marks,draw=black,fill=black] coordinates {
            (-1.3, 0, -0.5) 
            (-0.5,    0, -1) 
    	    };
            \draw [fill=gray!20,opacity=0.4]
            (-2.7,0,-1.2) -- (-0.7,0,-1.6) -- (0.3,0,0.3) -- (-1.7,0,0.7) -- (-2.7,0,-1.2);
            \draw [dashed]
            (-1.3, 0, -0.5) .. controls (-0.9,0,-0.7) ..  (-0.5,    0, -1);
           \draw [->, color = magenta] (-1.3, 0, -0.5) -- (-0.42, 0, -0.62);
           \draw [densely dotted] (-0.42, 0, -0.62) -- (-0.5,    0, -1);
        \node[font=\scriptsize, anchor=south west] at (0.15,0,0.15) {${T}_{\mathbf{w}} \mathbb{T}_2$};
        \node[font=\scriptsize] at (-1.5, 0, -0.3) {$\mathbf{w}$};
        \node[font=\scriptsize] at (+0.2, 0, -0.3) {\textcolor{magenta}{$\mathbf{u}=\log_{\mathbf{w}}(\tilde{\mathbf{w}})$}};
        \node[font=\scriptsize, anchor=north west] at (-0.55, 0,-0.8) {$\tilde{\mathbf{w}}=\exp_{\mathbf{w}}(\mathbf{u}) $};
        \end{axis}
        \end{tikzpicture}
        \captionof{figure}{Illustration of geometric tools on $\mathbb{T}_2$ (represented as a torus embedded in $\mathbb{R}^3$).}
        \label{fig:riem_torus}
    \end{minipage}
\end{figure}

Due to the specificity of SAR images (complex-valued data, speckle noise, etc.), recovering $\Delta\Phi_\text{disp}$ from actual measurements $\mathbf{w}$ remains a challenging issue that motivates the development of dedicated algorithms.
As mentioned previously, many approaches were investigated \cite{xu2020insar}, ranging from image processing \cite{song2017insar, deledalle2010nl}, to deep learning \cite{insarmonet_vitale_2022, insar2insar_geara_2024}, to multitemporal approaches \cite{guarnieri_2008, ferretti_2011, fornaro2014caesar, ning_cao_2015, Ansari2018}.
%
%
Multitemporal methods process a stack of $T$ co-registered phase images, which leads to $T(T-1)/2$ possible interferograms, denoted $\mathbf{w}_{ij},~\forall (i,j)\in[\![1,T ]\!]^2,$ and with $i>j$.
In this context, interferometric phase linking refers to a technique that produces estimates of each interferogram that satisfy the phase closure property:
\begin{equation*}
    \Delta\Phi_\text{disp}^{ij} + \Delta\Phi_\text{disp}^{jk} + \Delta\Phi_\text{disp}^{ki}
    =     \mathbf{0}  
\end{equation*}
for all image triplets with index $(i,j,k)\in[\![1,T ]\!]^3$.
Notice that denoising each interferogram individually would, in general, not yield such a relationship. 
In this paper, we will rely on the algorithm \texttt{COFI-PL} \cite{vu2025covariance} that achieves this by fitting the aforementioned phase structure to the estimated covariance matrix of local ($T$-dimensional) pixel patches. 
This phase linking process is done for all pixels over a sliding window. While highly effective, such approaches require access to the complete image timeseries and can therefore only be applied once all acquisitions have been collected.

\subsection{Interferograms as points on a Riemannian manifold}
\label{sec:phase_manifold}

Interestingly, interferograms exhibit a $2\pi$-periodicity, implying that the value of each pixel ranges from $0$ to $2\pi$ before being wrapped back to $0$. This periodicity constitutes strong prior knowledge regarding the geometric structure of the manifold on which the data resides. As illustrated in \cref{fig:datamanifold}, a one-pixel interferogram corresponds to a scalar value in the interval $[0, 2\pi)$ and can therefore be naturally associated with a point on the one-dimensional circle $\mathbb{S}_1$. Extending this concept, a two-pixel interferogram can be represented as a point on the two-dimensional flat torus $\mathbb{T}_2$. More generally, interferograms of dimension $d = w \times h$, corresponding to images of width $w$ and height $h$, can be interpreted as points on the $d$-dimensional flat torus $\mathbb{T}_d$.
Thus, we possess strong prior knowledge of the data manifold, which encodes the cyclical nature of interferograms. A data point on this manifold is \textit{valid} with respect to the cyclical constraint.

The $d$-dimensional torus $\mathbb{T}_d$ is a compact smooth manifold embedded in $\mathbb{R}^d$.
This allows us to leverage tools from differential geometry to properly handle points (and relationship between them) in this space.
Detailed introductions to these concepts can be found in \cite{absil2008optimization, boumal2023introduction, bouchard2024fisher}.
We review here some basic tools that will help in following the rest of the paper.
We also note that some tools can intuitively be recovered by identifying $\mathbb{T}_d$ to a product of $d$ circles (each circle representing the space for an entry of $\mathbf{w}$).
The tangent space at point $\mathbf{w}$ is denoted $T_{\mathbf{w}} \mathbb{T}_d$, and can be mapped to the space $\mathbb{R}^d$.
Here, we consider $\mathbb{T}_d$ as a Riemannian sub-manifold, which means that we endow each tangent space with the Euclidean metric.
This defines an inner product on each space $T_{\mathbf{w}}\mathbb{T}_d$, leading to the norm $\lVert \mathbf{u} \rVert_g^2 = \langle \mathbf{u}, \mathbf{u} \rangle$, for each tangent vector $ \mathbf{u} \in T_{\mathbf{w}}\mathbb{T}_d$.
This choice yields a geometry for $\mathbb{T}_d$, notably inducing the notion of \textit{geodesics}.
In particular, this provides tools such as the Riemannian exponential and Riemannian logarithm mappings:

\begin{equation}
    \log_{\mathbf{w}} : \mathbb{T}_d \to T_{\mathbf{w}}\mathbb{T}_d, \qquad \exp_{\mathbf{w}} : T_{\mathbf{w}}\mathbb{T}_d \to \mathbb{T}_d.
\label{eq:logexpspaces}
\end{equation}
The logarithm mapping $\log_{\mathbf{w}}(\cdot)$ can intuitively be understood as follows: given a point $\tilde{\mathbf{w}}$, $\mathbf{u}= \log_{\mathbf{w}}(\tilde{\mathbf{w}})$ is a vector in the tangent space of $\mathbf{w}$ that ``points towards'' $\tilde{\mathbf{w}}$, and whose norm  $\lVert \mathbf{u} \rVert_g^2$ is the \textit{geodesic distance} between the two points. This mapping is central to the definition of velocity vectors in the Riemannian flow matching framework (\cref{subsec:rie_fm}).
The exponential mapping $\exp_{\mathbf{w}}(\cdot)$ defines the inverse of the logarithm.
These tools are illustrated in \cref{fig:riem_torus}, and their expression will be defined in \cref{eq:log_exp}.

\section{A primer on Euclidean flow matching}

\subsection{Flow Matching}
The goal of flow matching is to learn a time-dependent mapping $\phi : [0,1] \times \mathbb{R}^d \rightarrow \mathbb{R}^d$,
called the \textit{flow}, which transports data points sampled from the initial distribution $p_0$ to intermediate distributions $p_t$ over time. We denote by $\phi_t(x_0)$ the position at time $t$ of a particle that started at $x_0$. From a probabilistic perspective, the distribution $p_t$ corresponds to the distribution of transported particles and can be written as:
\begin{equation}
p_t = [\phi_t]_{\#} p_0,
\end{equation}
where $[\phi_t]_{\#}$ denotes the push-forward operator acting on probability measures~\cite{santambrogio2015optimal}.
Rather than learning the flow $\phi_t$ directly, the flow matching framework parametrizes it through a time-dependent velocity field. We define the velocity field:
\[
v : [0,1] \times \mathbb{R}^d \rightarrow \mathbb{R}^d,
\]
which specifies, for every position $x \in \mathbb{R}^d$ and time $t \in [0,1]$, the instantaneous direction and speed of particles evolving under the flow. The flow $\phi_t$ is then induced by this velocity field through the following ordinary differential equation:

\begin{equation}
\begin{cases}
  \dfrac{d}{dt}\phi_t(x) = v_t(\phi_t(x)), \\[6pt]
  \phi_0(x) = x~.
\end{cases}
\label{eq:insarflow_flow}
\end{equation}

This ordinary differential equation defines how particles evolve under the velocity field $v_t$: starting from their initial position $x$ at time $t=0$, they follow the trajectories induced by $v_t$ and are transported toward the target distribution at $t=1$.

A natural training strategy is therefore to learn a neural network $v_\theta$, parameterized by $\theta$, that approximates the true (unknown) velocity field $v_t$. This leads to the following regression objective:
\begin{equation}
    \mathcal{L}(\theta)
    =
    \mathbb{E}_{\substack{t\sim\mathcal{U}([0, 1]) \\ x_t \sim p_t(x_t)}}
    \Big[
        \lVert v_\theta(t, x_t) - v_t(x_t) \rVert^2
    \Big],
\label{eq:fm_intractable_loss}
\end{equation}
where $\mathbb{E}$ denotes expectation, $\mathbb{E}_p[f] = \int f dP$, i.e \cref{eq:fm_intractable_loss} reads as:
$$
    \mathcal{L}(\theta)
    = \int_0^1 \int_\mathbb{{R}^d}
    \Big[
        \lVert v_\theta(t, x_t) - v_t(x_t) \rVert^2
    \Big]p_t(x_t)\text{d}x_t\text{d}t.
$$


If this objective were minimized exactly, the flow induced by the learned velocity field $v_\theta$ would define a valid transport between the distributions $p_0$ and $p_1$. However, this objective is not directly tractable in practice. Indeed, the intermediate distributions $p_t$ are not explicitly defined; many choices of $p_t$ can satisfy the flow equation \cref{eq:insarflow_flow}, and the true velocity field $v_t$ is unknown and does not admit a closed-form expression.

To overcome this difficulty, flow matching replaces the regression on true velocities with a regression on \textit{conditional velocities}, which can be computed analytically under suitable assumptions \cite{lipman_flow_2022, albergo_stochastic_2023}.

\subsection{Interpolant}

Flow matching models construct the intermediate points $x_t$, and implicitly the distributions $p_t$, by interpolating between samples drawn from the endpoint distributions $p_0$ and $p_1$. 
The interpolants are parameterized by time~$t$, and a simple yet commonly used choice is the linear interpolant~$I_t$:
\begin{equation}
    x_t = I_t(x_0, x_1) = (1 - t) x_0 + t x_1.
\label{eq:linear_interpolant}
\end{equation}
The linear interpolations correspond to \textit{geodesics in Euclidean space}. 
However, this is not valid for all manifolds and we introduce its adaptation to Riemannian manifolds in \cref{subsec:rie_fm}.
For example, on a sphere, the shortest path between two points follows a great-circle arc rather than a straight line through the interior. 

\subsection{Coupling}

Sampling from the conditional distribution $p(x_t \mid x_0, x_1)$ then amounts to first sampling a pair $(x_0, x_1)$ from a joint distribution $p(x_0, x_1)$ referred to as a \textit{coupling}, and subsequently computing the interpolated point using \cref{eq:linear_interpolant}. The simplest choice is the \textit{independent coupling}, where $x_0$ and $x_1$ are sampled independently, i.e., $p(x_0, x_1) = p(x_0)p(x_1)$. More sophisticated couplings have been proposed to use additional structure or prior knowledge~\cite{albergo2023couplings, tong_simulation-free_2023-1}.

\subsection{Conditional velocity}

In the case of linear interpolants \cref{eq:linear_interpolant}, the \textit{conditional velocity} admits a closed-form expression. It is given by the time derivative of the interpolation path connecting the two endpoints $x_0$ and $x_1$:
\begin{equation}
    v_t(x_t \mid x_0, x_1)
    =
    \frac{d}{dt} I_t(x_0, x_1)
    =
    \frac{d}{dt} \big[(1 - t) x_0 + t x_1\big]
    =
    x_1 - x_0.
\label{eq:conditional_velocity}
\end{equation}
This expression highlights that, conditionally on $(x_0, x_1)$, the induced trajectory is a straight line with constant velocity. 

\subsection{Loss function}

As discussed above, directly regressing the marginal (or true) velocity field $v_t(x_t)$ is intractable (see \cref{eq:fm_intractable_loss}), since neither the intermediate distribution $p_t$ nor the true velocity field are available in closed form. However, Lipman et al.~\cite{lipman2022flow} show that minimizing a regression objective on conditional velocities is sufficient: any minimizer of the conditional velocity regression problem is also a minimizer of the marginal velocity regression problem.

Consequently, the \textit{conditional flow matching} objective used to train a neural network $v_\theta$ to approximate the marginal velocity field is defined as:

\begin{equation}
    \mathcal{L}_{\text{FM}}(\theta) = \mathbb{E}_{\substack{t\sim\mathcal{U}([0, 1]) \\ (x_0, x_1) \sim p(x_0, x_1) \\ x_t\sim p_t(x_t\mid x_0, x_1)}}\Bigg[\Big\lVert v_\theta(t, x_t) - v_t(x_t\mid x_0, x_1) \Big\rVert^2\Bigg]~.
\label{eq:fm_loss}
\end{equation}
For linear interpolants, this objective reduces to:
\begin{equation}
    \mathcal{L}_\text{FM}(\theta) = \mathbb{E}_{\substack{t\sim\mathcal{U}([0, 1]) \\ (x_0, x_1) \sim p(x_0, x_1) \\ x_t\sim p_t(x_t\mid x_0, x_1)}}\Bigg[\Big\lVert v_\theta(t, x_t) - (x_1 - x_0) \Big\rVert^2\Bigg]~.
\label{eq:fm_loss_linear}
\end{equation}

\subsection{Sampling} At inference time, points $x_0 \sim p_0$ are transported to the target distribution $p_1$ by following the learned velocity field:
\begin{equation}
    \hat{x}_1 = \varphi_{t=1}(x_0)
    = \text{ODE}^{v_\theta}(x_0, 0 \rightarrow 1),
\label{eq:insarflow_ode}
\end{equation}
where $\text{ODE}^{v_\theta}(\cdot, t_0 \rightarrow t_1)$ denotes solving \cref{eq:insarflow_flow} between times $t_0$ and $t_1$ with a numerical solver, replacing the velocity with its neural approximation $v_\theta(t, \cdot)$.

\section{Riemannian Flow Matching for InSAR}
\label{subsec:rie_fm}

\begin{figure*}[t!]
\centering
\includegraphics[width=0.8\linewidth, trim=0 0 50 0]{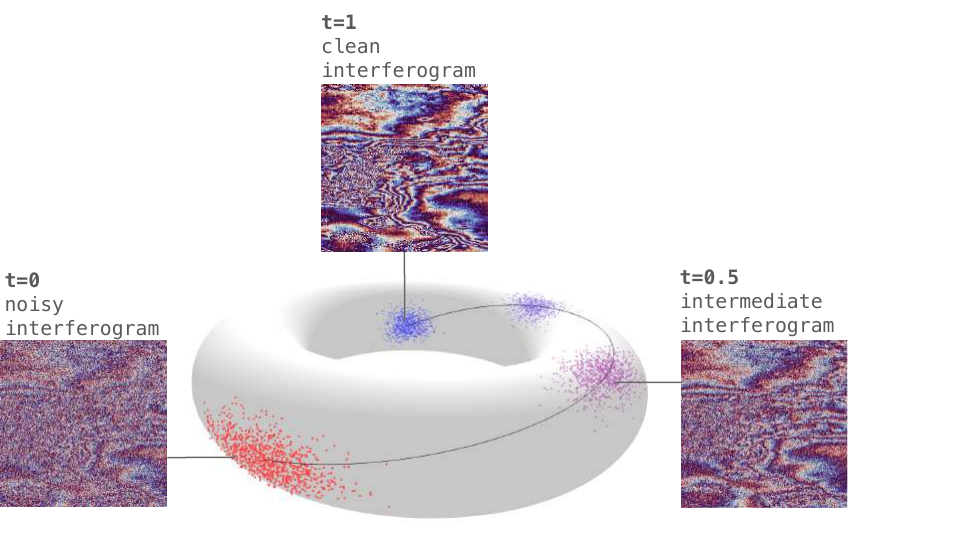}
\caption[InSARFlow generates denoised versions $\varphi_1(\mathbf{w}_0)$ of two interferograms $\mathbf{w}_0$ corresponding to two different temporal gaps $\Delta t$.]{\insarflow{} framework: noisy interferograms at $t=0$ are progressively transported toward clean interferograms at $t=1$. The method models interferograms as points on a flat torus, ensuring consistency with phase periodicity throughout the generative process.}
\label{fig:torus_denoising}
\end{figure*}

The use of flow models in the InSAR domain requires several adaptations to account for the underlying geometry of the data.

As discussed in \cref{sec:phase_manifold}, each pixel of an interferogram encodes a phase $\phi \in [0, 2\pi)$. 
Consequently, an interferogram with $d$ pixels can be represented as a vector of phases 
$\mathbf{w} = [\phi_0, \phi_1, \ldots, \phi_{d-1}]$, which corresponds to a point on the Riemannian torus 
$\mathbb{T}_d \sim (S^1)^d$. 
To ensure the validity of generated interferograms, we employ a generative model that operates directly on this Riemannian phase manifold.

\subsection{Training}

{
\begin{algorithm}
\caption{\insarflow{} Training Process}
\label{alg:training}
\begin{algorithmic}[1]

\Require Paired dataset $\{(\mathbf{w}_0, \mathbf{w}_1)_i\}_{i \in \llbracket1, N\rrbracket}$
\State Initialize parameters $\theta$ of $v_\theta$
\While{not converged}
    \State sample time $t \sim \mathcal{U}(0, 1)$
    \State sample training pair from dataset $(\mathbf{w}_0, \mathbf{w}_1) 
    \sim p(\mathbf{w}_1 \mid \mathbf{w}_0)p(\mathbf{w}_0)$
    \State compute interpolant $\mathbf{w}_t = \exp_{\mathbf{w}_0}(\kappa(t) \log_{\mathbf{w}_0}(\mathbf{w}_1))$
    \State $\mathcal{L}(\theta) = \|v_\theta(t, \mathbf{w}_t) - v_t(\mathbf{w}_t\mid \mathbf{w}_0, \mathbf{w}_1)\|_g^2$
    \State $\theta = \texttt{optimizer\_step}(\mathcal{L}(\theta))$
\EndWhile
\State \textbf{return} $v_\theta(t, \mathbf{w}_t)$
\Comment{trained velocity estimator}
\end{algorithmic}
\end{algorithm}
}

The generalization of flow matching models to Riemannian geometries was proposed by Chen and Lipman~\cite{chen2024flow}.
The training objective in \cref{eq:fm_loss} can be generalized by observing that, at each point 
$\mathbf{w}_t$ on the Riemannian manifold, both the estimated velocity 
$\mathbf{v}_\theta(t, \mathbf{w}_t)$ and the true conditional velocity 
$\mathbf{u}_t(\mathbf{w}_t \mid \mathbf{w}_0, \mathbf{w}_1)$ lie in the tangent space of the torus at $\mathbf{w}_t$. 
We denote this tangent space by $T_{\mathbf{w}_t}\mathbb{T}_d$ (as illustrated in \cref{fig:riem_torus}). The training process is detailed in \cref{alg:training}.

Here, we consider $\mathbb{T}_d$ as a submanifold and thus use the induced Euclidean metric on the space $T_{\mathbf{w}_t}\mathbb{T}_d$, leading to the standard Euclidean norm $\lVert u \rVert_g^2 = \langle u, u \rangle$.
The objective (\cref{eq:insarflow_flow}) becomes:
\begin{equation}
    \mathcal{L}_\text{RFM}(\theta)
    =
    \mathbb{E}_{\substack{\mathbf{w}_0, \mathbf{w}_1 \sim p(\mathbf{w}_0, \mathbf{w}_1)\\ t \sim U([0, 1])}}
    \big\lVert
        \mathbf{v}_\theta(t, \mathbf{w}_t)
        -
        \mathbf{v}_t(\mathbf{w}_t \mid \mathbf{w}_0, \mathbf{w}_1)
    \big\rVert_g^2.
\label{eq:rfmloss}
\end{equation}
 The endpoints are sampled as before from the coupling $p(\mathbf{w}_0, \mathbf{w}_1)$, but the chosen interpolant must respect the geometry of the manifold so that all constructed points $\mathbf{w}_t$ belong to the torus $\mathbb{T}_d$. In other words, the interpolation paths must follow the \textit{geodesics on the Riemannian manifold}. Thus, we use the following interpolant:
\begin{equation}
    \mathbf{w}_t = I_t(\mathbf{w}_0, \mathbf{w}_1)
    = \exp_{\mathbf{w}_1}\!\big(\kappa(t)\, \log_{\mathbf{w}_1}(\mathbf{w}_0)\big)
\label{eq:insarflow_riem_inter}
\end{equation}
where $\kappa$ is a monotone decreasing, differentiable function satisfying $\kappa(0)=1$ and $\kappa(1)=0$, enabling interpolation between $p_0$ and $p_1$. For this interpolant, the conditional velocity is ${\mathbf{v}_t}(\mathbf{w}_t \mid \mathbf{w}_0, \mathbf{w}_1) = \dot{\kappa}(t)\log_{\mathbf{w}_1}(\mathbf{w}_0)$.
As presented in \cref{fig:riem_torus} and \cref{eq:logexpspaces}, $\exp_{\mathbf{w}}$ maps the tangent space to the manifold and $\log_{\mathbf{w}}$ maps the manifold to its tangent space~\cite{boumal2023introduction}. 
The interpolation defined in \cref{eq:insarflow_riem_inter} can be understood as a three-step procedure:
\begin{enumerate}
    \item Compute the tangent vector $\log_{\mathbf{w}}(\mathbf{w}_0) \in T_{\mathbf{w}}\mathbb{T}_d$, which points from $\mathbf{w}_1$ toward $\mathbf{w}_0$ and whose norm equals the geodesic distance between $\mathbf{w}_1$ and $\mathbf{w}_0$.
    \item Perform a linear interpolation in the tangent space $T_{\mathbf{w}}\mathbb{T}_d$ by scaling this vector with the time-dependent factor $\kappa(t)$.
    \item Map the resulting point back onto the Riemannian manifold using the exponential map $\exp_{\mathbf{w}_1}$.
\end{enumerate}

We know that the interferograms lie on the flat $d$-dimensional torus $[0, 2\pi)^d$. The commonly used maps for this manifold $\exp_{\mathbf{w}}(\cdot)$ and $\log_{\mathbf{w}}(\cdot)$ (applied elementwise to each entry) are:
\begin{equation}
\left\{
    \begin{aligned}
        \exp_w(u) &= (w + u) \pmod{2\pi} \\
        \log_{w_a}(w_b) &= \text{arctan2}\big(\sin(w_b - w_a),\, \cos(w_b - w_a)\big)
    \end{aligned}
\right.
\label{eq:log_exp}
\end{equation}

{
\subsection{Sampling}
\begin{algorithm}[t]
\caption{\insarflow{} Sampling Process}
\begin{algorithmic}[1]

\Require Trained model $v_\theta$, noisy observation $\mathbf{w}_0$, number of steps $S$
\State Set step size $\Delta t = 1/S$
\State Initialize $\mathbf{w}_t \leftarrow \mathbf{w}_0$
\For{$s = 0, 1, \dots, S-1$}
    \State $t \leftarrow s \cdot \Delta t$
    \State Predict velocity $\hat{v} \leftarrow v_\theta(t,\, \mathbf{w}_t)$
    \State Euler step on $\mathbb{T}_d$: $\quad\mathbf{w}_{t+\Delta t} = \exp_{\mathbf{w}_t}\!\big(\Delta t \cdot \hat{v}\big)$
\EndFor
\State \Return $\mathbf{w}_1$ \Comment{denoised interferometric phase}
\end{algorithmic}
\label{alg:sampling}
\end{algorithm}
}
To generate new data points with Riemannian flow matching models, the flow equation in \cref{eq:insarflow_flow} must be solved directly on the manifold.
We rely on Riemannian ODE solvers, which guarantee that the generated samples $\mathbf{w}_t$ remain on the flat torus at every integration step~\cite{chen2024flow}. We detail the sampling process in \cref{alg:sampling}. 
A simplified example of transport between two distributions $p_0$ and $p_1$ on the torus $\mathbb{T}_2$ is illustrated in \cref{fig:torus_denoising} (top).

{
\section{Experimental protocol}
\subsection{Datasets}

We evaluate \insarflow{} on two distinct datasets. First, we observe the denoising performances of \insarflow{} in the case of real-world applications on the Mexico City dataset (\cref{sub:real_dataset}). Nevertheless, the absence of ground truth for real datasets makes fair comparison to other baselines difficult. Then, we conducted evaluation on a synthetic dataset (\cref{sub:synth_dataset}).

\subsubsection{Real-world dataset}
\label{sub:real_dataset}

We use real data collected over Mexico City.
We have a data cube containing \num{40} Sentinel-1 SLC SAR acquisitions of size $3638\times16709$ pixels at \num{10}m GSD, acquired every 12 days between August 14, 2019 and December 6, 2020~\cite{vu2023robust, elhajjar:hal-04770259, 10858717, vu2025covariance}, preprocessed using SNAP.
The acquisitions are processed so as to keep only the phase differences between each acquisition time $T^\text{acq}_i$ and the initial date $T^\text{acq}_0$. To build (noisy, clean) interferogram training pairs, we leverage \texttt{COFI-PL}~\cite{vu2025covariance}, a \textbf{multi-temporal} denoising algorithm. Given a noisy interferogram timeseries, \texttt{COFI-PL} produces the corresponding cleaned timeseries. The training pairs are composed of a single noisy interferogram and its cleaned version from the \texttt{COFI-PL} predicted timeseries. To ensure sufficient spatial context while maintaining reasonable compute costs, the interferograms are downsampled by decimation with factor \num{4} and cropped into phase-difference images of size \num{256}$\times$\num{256}, with \num{75}\% overlap for training, resulting in a total of \num{34 747}  training examples.

For evaluation purpose, we split the dataset into two spatially disjoint areas, avoiding spatial information leakage. The validation and test subsets are respectively of size \num{100} and \num{741}.

\subsubsection{Simulated dataset}
\label{sub:synth_dataset}
Synthetic interferograms of size \num{256}$\times$\num{256} are generated with an SNR uniformly sampled in $[0.2, 0.5]$. The simulator randomly combines atmospheric turbulence, ground deformation signals, urban structures, and phase discontinuities. Acquisition geometry and spatial resolution match Sentinel-1 characteristics. DEMs are randomly extracted from the RGE ALTI \footnote{RGE ALTI:\href{https://www.data.gouv.fr/datasets/rge-alti-r}{https://www.data.gouv.fr/datasets/rge-alti-r}} at 10m resolution.
Further details on the simulation are provided in~\cite{insar_sim, insar_sim2}.
The dataset has a training set of \num{184 921} images and validation and test sets of respective sizes \num{100} and \num{250}.
}

{

\subsection{Applications}

\begin{table}[t]
\centering
\caption{\insarflow{} application to interferogram generation and denoising. While the two applications use a transport towards $p_1$, the distribution of clean interferograms, they use different initial distributions $p_0$ and couplings.}
\label{tab:applications}
{\footnotesize
{\small\noindent
\begin{tabularx}{\linewidth}{@{}lXX@{}}
\toprule
& Generation & Denoising\\
\midrule
Initial distribution & Noise & Noisy interferograms\\
$p_0$ & $p_0=\exp_{0_d}\mathcal{N}(0, I_d)$ & $p_0 = p_\text{noisy}$\\
\midrule
Final distribution & Clean interferograms & Clean interferograms\\
$p_1$ & $p_1 = p_\text{clean}$ & $p_1 = p_\text{clean}$\\
\midrule
Coupling & Independent & Data-dependent\\
$p(\textbf{w}_0, \textbf{w}_1)$ & $p(\textbf{w}_0)p(\textbf{w}_1)$ & $p(\textbf{w}_1 \mid \textbf{w}_0)p(\textbf{w}_0)$\\
\bottomrule
\end{tabularx}
}

}
\end{table}

In this section we study the use of \insarflow{} for two separate applications: first, in \cref{sub:generation}, we generate new clean interferograms and use Riemannian Flow Matching as an interferogram simulator. Second, in \cref{sub:denoising}, we denoise noisy interferograms with \insarflow{}. We summarize the main differences from the transport perspective in \cref{tab:applications}.

\subsubsection{Generation}
\label{sub:generation}
}

Flow models can be used as generative models by transporting a simple, known prior $p_0 = p_{\text{prior}}$ that we can sample from to the data distribution $p_1 = p_{\text{data}}$.
We train \insarflow{}, a Riemannian flow matching model to generate phase differences by transporting a prior distribution toward the Mexico data cube denoised with the \texttt{COFI-PL} algorithm~\cite{vu2025covariance}.

Concretely, we use the prior $p_0 = \exp_{0_d}\mathcal{N}(0,I_d)$ and the data distribution $p_1 = p_{\text{clean}}$ consisting of the clean interferograms.
%


\subsubsection{Denoising}
\label{sub:denoising}

Flow models enable transport between two arbitrary distributions $p_0$ and $p_1$, thereby opening the door to new denoising strategies, with transports between a distribution of noisy data and its clean counterpart.

The main strength of flow-based methods is to learn a transport only by sampling interpolants between noisy and clean interferograms. They do no not require explicit assumptions about the nature of the corruption noise, nor does it require knowledge of how to simulate the corruption process itself. In contrast, Gaussian diffusion models typically assume that the corruption noise follows a Gaussian distribution.

The interferograms in the datasets under study exhibit significant noise, making them challenging to use for accurate displacement or ground deformation estimation, thus motivating the need for a dedicated denoising step. Therefore, we train \insarflow{} for the denoising task on two distinct datasets:

\begin{figure}[t]
\centering
\includegraphics[width=\linewidth]{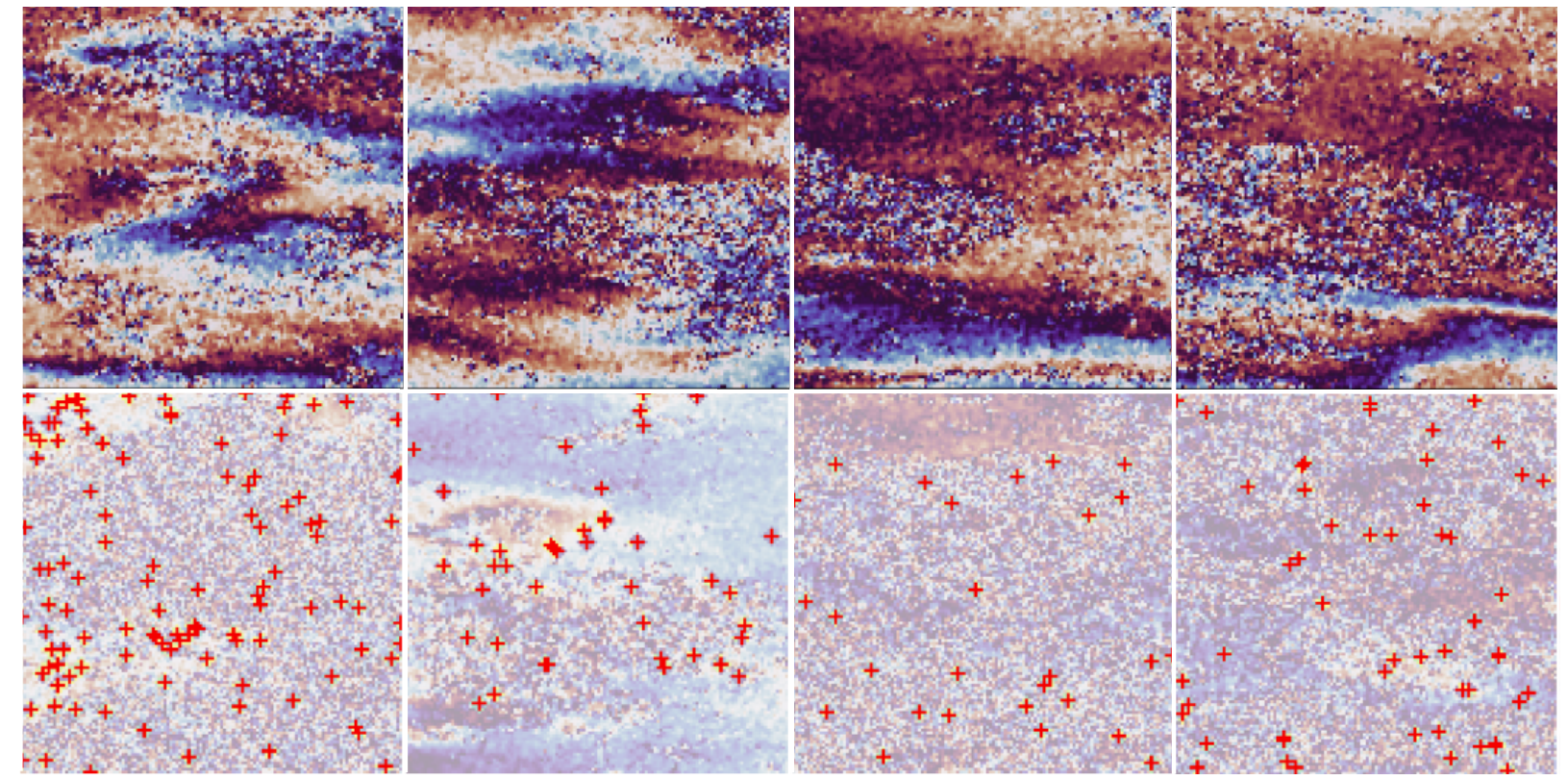}
\caption[\textbf{(Top)} Tiles generated by \insarflow{} on the Riemannian manifold $\mathbb{T}_d$ using 25 integration steps. By construction, all generated pixels are valid, \ie all phase values lie within the interval $[0, 2\pi)$.
\textbf{(Bottom)} Tiles generated by the flow-based model on the Euclidean manifold $\mathbb{R}^d$ using 25 integration steps.]{\textbf{(Top)} Tiles generated by \insarflow{} on the Riemannian manifold $\mathbb{T}_d$ using 25 integration steps. By construction, all generated pixels are valid, \ie all phase values lie within the interval $[0, 2\pi)$.
\textbf{(Bottom)} Tiles generated by the flow-based model on the Euclidean manifold $\mathbb{R}^d$ using 25 integration steps.}
\label{fig:generation-euc-riem}
\end{figure}

\begin{figure*}[t]
\centering
\input{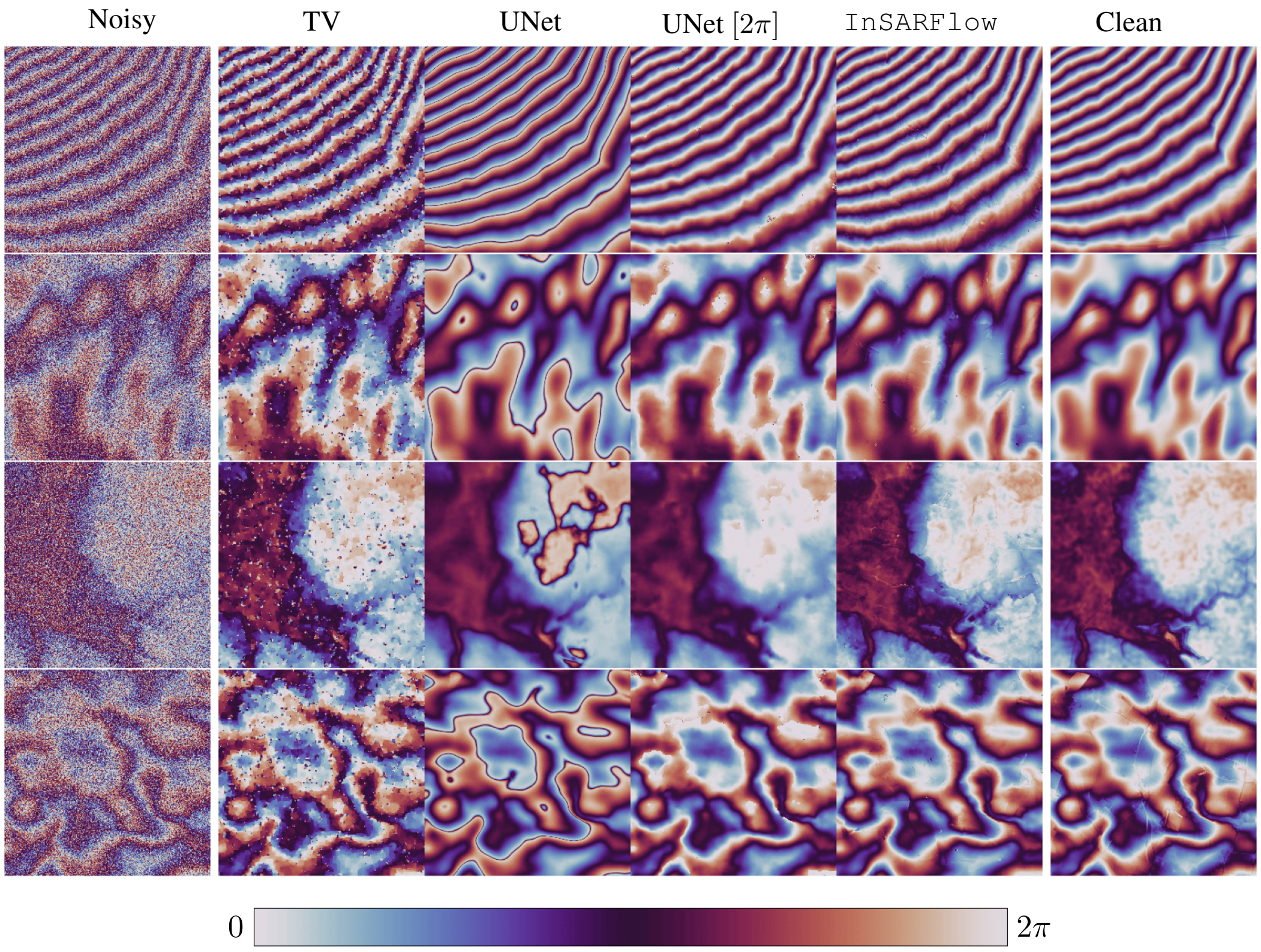}
\caption{Comparison of denoising results. From left to right: noisy input, total variation denoiser (TV), U-Net, U-Net $[2\pi]$, \insarflow{}, clean ground truth. \insarflow{} generates interferograms with high-frequency details and without visible artifacts.}
\label{fig:main_denoised_synthetic}
\end{figure*}

{
\begin{enumerate}
    \item \textbf{Synthethic dataset:} synthetic dataset, we learn a transport between the noisy interferograms and the ground truth clean interferograms. On this scenario, we have access to the ground truth clean interferograms as they are generated by the simulator alongside the noisy interferograms.

    \item \textbf{Mexico City dataset:} we learn the flow between the distribution of real phase differences and the distribution of phase differences denoised using the \texttt{COFI-PL} algorithm~\cite{vu2025covariance}. 
    This way, this procedure has the advantage of distilling the \textit{a priori} knowledge about the spatial and temporal structure of the noise removed by \texttt{COFI-PL}, a multitemporal denoising algorithm, into \insarflow{}, a single temporal model. This allows us to assess how well the Riemannian-flow-based denoiser can reproduce the quality of a multitemporal denoising method using only a single interferogram (instead of a full timeseries).
\end{enumerate}
}

{
\subsection{Implementation details}

\subsubsection{Neural network backbone}
\label{sub:backbone}

To ensure fair comparison, \insarflow{} and other neural-based baselines share the same convolutional backbone, namely the \texttt{FCDM} backbone \cite{kwon2026reviving}. We use the official implementation \footnote{Official implementation:\href{https://github.com/star-kwon/FCDM}{ https://github.com/star-kwon/FCDM}}, the backbone of size B ($\approx$\num{126}M parameters) with hidden size=256,  and depth values of [2,4,8,4,2]. For non-time-dependent methods, e.g. supervised UNets, the existing time-conditioning is deactivated. 
All models are trained for trained for \num{100000} iterations with batches of \num{32} examples. We use a learning rate of \num{1e-4}, gradient clipping at \num{0.5}.

\subsubsection{\insarflow{} implementation}

We approximate the velocity $v_\theta(t, \textbf{w}_t)$ with the above mentioned backbone (\cref{sub:backbone}). We adopt the common practice of leveraging the exponential moving average (EMA) of weights to stabilize training. At inference, we generate clean interferograms $\hat{\textbf{w}}_1 \sim p_1$ by integrating the ODE \cref{eq:insarflow_flow} using an explicit Euler solver (see \cref{alg:sampling}) with 10 discretization steps, i.e. 10 NFE (number of function evaluation).

\subsubsection{Denoising baselines implementations}

\textbf{Supervised UNets}
To evaluate the advantages of generative denoising proposed in \insarflow{} compared to supervised denoising, we train two supervised denoisers with the same backbone as \insarflow{} (see \cref{sub:backbone}). The first one, referred as UNet, is trained in the euclidean space with standard MSE loss. The second, denoted as UNet $[2\pi]$, is trained with a cyclical loss function $\mathcal{L}(\textbf{w}_0;\psi) = 1 - \cos(\textbf{w}_1 - \hat{\textbf{w}}_1)$, where $\hat{\textbf{w}}_1$ is the prediction of UNet $[2\pi]$ of weights $\psi$. Thus, this model takes into account the cyclical nature of phase differences, i.e., the geometry of the manifold.

\textbf{TV denoiser of flat torus} We compare \insarflow{} to the \textit{Weinmann-Demaret-Storath} denoising algorithm \cite{tv_manifold}. It proposes a \textit{total-variation} (TV) regularization designed for manifold-valued images, with applications to InSAR denoising. We reimplemented their parallel proximal point algorithm (algorithm 2 in the original paper), and tuned the regularization factor $\alpha$ according to the circular MSE on samples from the validation set and report the evaluations metrics on the test set for this optimal value.

\subsubsection{Evaluation metrics}

In the following, we evaluate all the models with metrics that take into account the manifold geometry:
\begin{enumerate}
    \item \textbf{Circular MSE}: denoted MSE $[2\pi]$, to compute the angle distance between the true clean interferogram $\textbf{w}_1$ and the predicted one $\hat{\textbf{w}}_1$.

\begin{equation}
    \text{CMSE}(\hat{\textbf{w}}_1, {\textbf{w}}_1) = \frac{1}{N}\sum_{k=1}^d \left[\text{atan2}\!\left(\sin(\hat{\textbf{w}}_1^k - {\textbf{w}}_1^k),\ \cos(\hat{\textbf{w}}_1^k - {\textbf{w}}_1^k)\right)\right]^2~.
\end{equation}

    \item \textbf{Kullback-Leibler divergence}: denoted $D_\text{KL}$, for a pair of true and predicted interferograms ($\textbf{w}_1$, $\hat{\textbf{w}}_1$), we compute their respective pixel-values histrograms with $N_\text{bins}$ (denoted $\textbf{h}_1$, $\hat{\textbf{h}}_1$) and then the Kullback-Leibler divergence between the histograms:
\begin{equation}
    D_\text{KL}(\textbf{h}_1, \hat{\textbf{h}}_1) = \sum_{i=1}^{N_\text{bins}} \textbf{h}_1^{(i)}\log \frac{\textbf{h}_1^{(i)}}{\hat{\textbf{h}}_1^{(i)}}. \end{equation}
\end{enumerate}
}

\section{Results on generation task}


As illustrated by the qualitative examples in \cref{fig:generation-euc-riem} (top row), \insarflow{} generates plausible phase-difference images that exhibit the same characteristics as the real data: presence of fringes and salt-and-pepper noise. This approach could, for example, be used to simulate synthetic InSAR data, a technique commonly used in the literature to train and/or evaluate deep models~\cite{insarmonet_vitale_2022, insar2insar_geara_2024}.

Euclidean flow models fail to generate valid interferograms. 
Indeed, they produce samples that do not lie on the torus $\mathbb{T}_d$, with values falling outside the interval $[0, 2\pi)$, particularly near fringes, as illustrated in \cref{fig:generation-euc-riem} (bottom). These invalid samples would be problematic for subsequent phase unwrapping, and motivate the use of \insarflow{} for generation.


\section{Results on denoising}

\begin{figure*}[t!]
\centering
\input{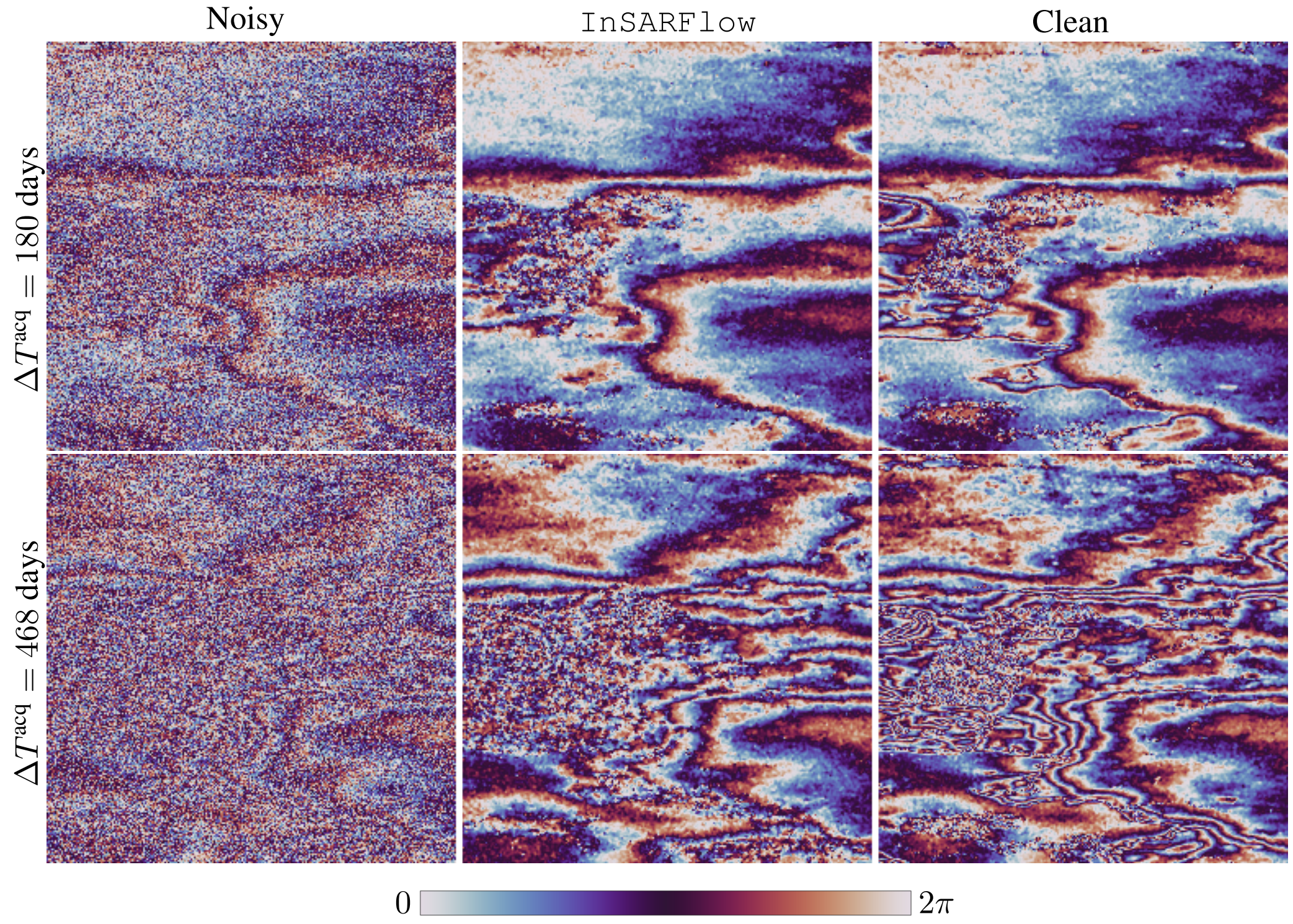}
\caption[InSARFlow generates denoised versions of two noisy interferograms corresponding to two different temporal gaps $\Delta T^\text{acq}$.]{InSARFlow generates denoised versions of two noisy interferograms corresponding to two different temporal gaps $\Delta T^\text{acq}$. Predictions obtained from a single interferogram using \insarflow{} are close to the clean interferograms produced by \texttt{COFI-PL}, a multi-temporal denoising algorithm, exploiting phase timeseries to form its estimates.}
\label{fig:main_denoised}
\end{figure*}

\begin{table}[b]
\centering
{
\caption{\textbf{Quantitative Results} on synthetic dataset. We report the circular MSE (MSE $[2\pi]$) and the $D_\text{KL}$ between pixel values histograms. \insarflow{} produces high-quality interferograms, closer to the ground truth clean interferograms according to both evaluation metrics.}}
\label{tab:synth_metrics}
{
\begin{tabular}{l cc}
        & \multicolumn{2}{c}{\textbf{Synthetic dataset}}\\
        
        \textbf{Method}& MSE $[2\pi]$ $\downarrow$& $D_\text{KL}$ $\downarrow$\\
        \midrule
        TV denoiser& 0.219& 0.038\\
        UNet& 0.154& 0.611\\
        UNet $[2\pi]$& 0.034
& 0.033\\

\midrule
 Flow Matching (NFE=10)& &\\
 Euclidean& 0.029&0.014\\
        \rowcolor{customblue!50} \insarflow& \textbf{0.026}& \textbf{0.013}\\
        \bottomrule
\end{tabular}
}
\end{table}

\subsection{Synthetic dataset}

First, \cref{fig:main_denoised_synthetic} presents the denoised interferograms produced by \insarflow{} and the different baseline methods.
The outputs generated by \insarflow{} are visually closer to the ground truth than those obtained with the other baselines.
The total-variation-based denoiser (TV) produces clean interferograms with noticeable artifacts.
The UNet trained with the Euclidean mean squared error loss generates blurry interferograms and exhibits visible artifacts.
The UNet $[2\pi]$, trained using a circular loss function, outperforms the standard UNet, although it still suffers from oversmoothing effects.
In contrast, \insarflow{} generates interferograms with high-frequency details.

The qualitative observations are supported by the denoising metrics reported in \cref{tab:synth_metrics}.
\insarflow{} achieves the best overall performance, outperforming the strongest non-generative baseline, UNet $[2\pi]$, on both evaluation metrics.
For the circular MSE, \insarflow{} attains a score of \num{0.026}, compared with \num{0.034} for UNet $[2\pi]$.
A similar trend is observed for the histogram divergence metric ($D_\text{KL}$), where \insarflow{} obtains \num{0.013} against \num{0.033} for UNet $[2\pi]$.
These results also highlight the advantage of Riemannian flow matching over standard Euclidean flow matching, as \insarflow{} achieves a lower circular MSE than its Euclidean counterpart (\num{0.026} versus \num{0.029}).

\begin{figure*}[t!]
\centering
\includegraphics[width=0.9\linewidth]{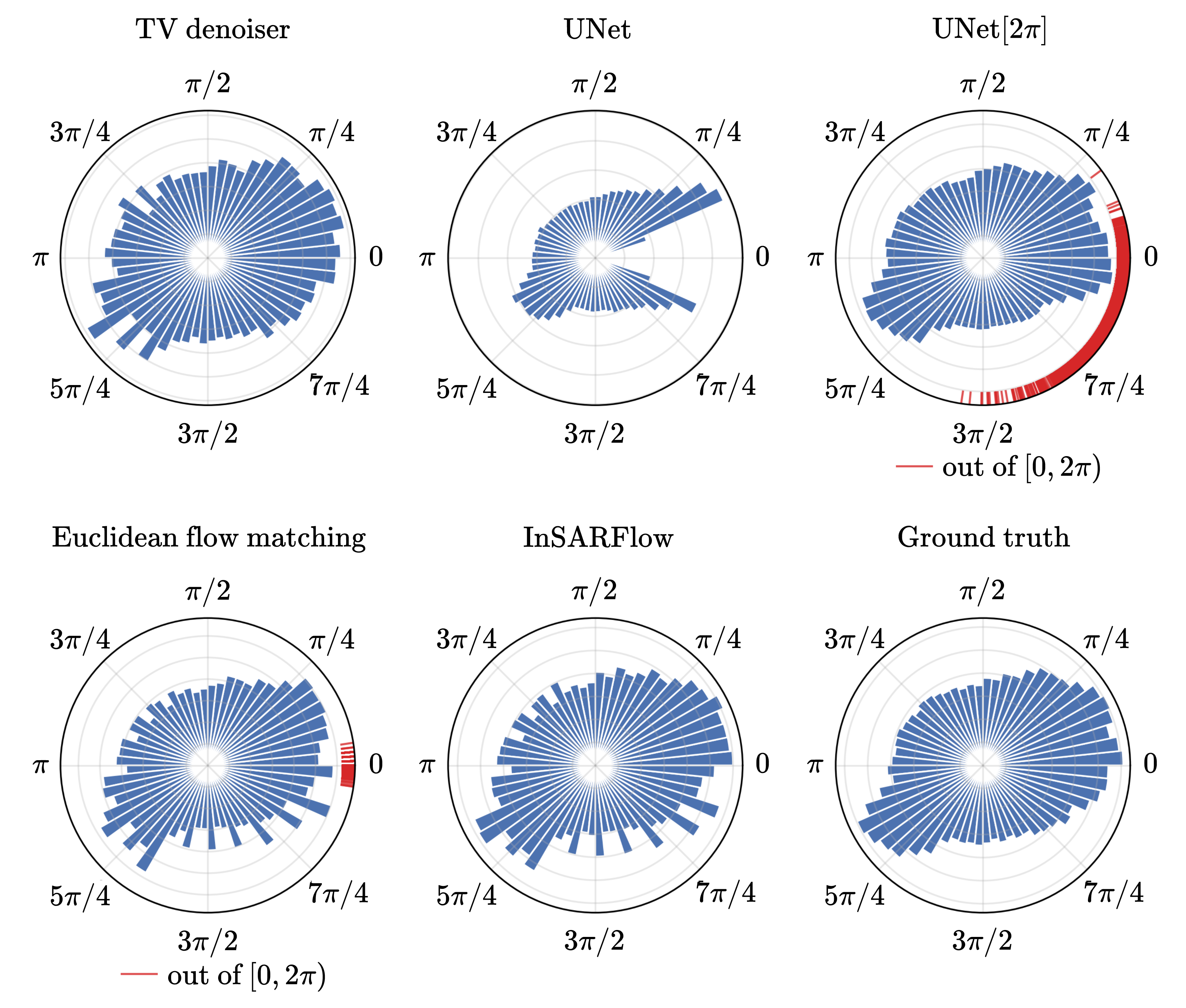}
\caption{Polar distribution of phase estimates produced by different denoising and reconstruction methods compared to the ground truth.}
\label{fig:circular_distribution}
\end{figure*}

In addition, we show in \cref{fig:circular_distribution} the circular histograms for all models.

\textbf{Histogram gaps}
Interestingly, the standard UNet baseline exhibits a noticeable gap in the distribution of pixel values around \num{0}.
This behavior can be explained by the use of a Euclidean mean squared error loss during training. For phase values close to \num{0} and $2\pi$, the reconstruction error can become artificially large despite a small angular discrepancy. For example, if the ground-truth value is $2\pi$ and the model predicts $\varepsilon$, the standard MSE assigns an error close to $2\pi-\varepsilon$, whereas the true angular error is only $\varepsilon$ when the periodic nature of the phase is taken into account.
We hypothesize that this excessive penalty near the phase discontinuity encourages the model to avoid predicting values in this region.
This issue is resolved when using a circular MSE loss: the UNet $[2\pi]$ model does not exhibit the gap around \num{0} observed in the pixel-value histograms of the standard UNet.

\textbf{Out-of-range pixel values}
Because the UNet $[2\pi]$ is trained with circular MSE, which does not penalize points outside $[0, 2\pi)$, the model produces such pixel values.
In this case, applying a mapping to ensure all pixel values are in $[0, 2\pi)$ is possible, since it reflects strategy adopted at training time.
As in the generation case, the Euclidean flow matching model also exhibits pixel-values outside $[0, 2\pi)$.
This time, these errors come from the velocity prediction error and discretization errors in the sampling process and are not due to a loss function that takes the circular nature of the data into account.

        


\subsection{Mexico City dataset}
\begin{figure}[t]
    \centering
    \begin{subfigure}[b]{0.49\linewidth}
        \centering
        \includegraphics[width=0.8\linewidth]{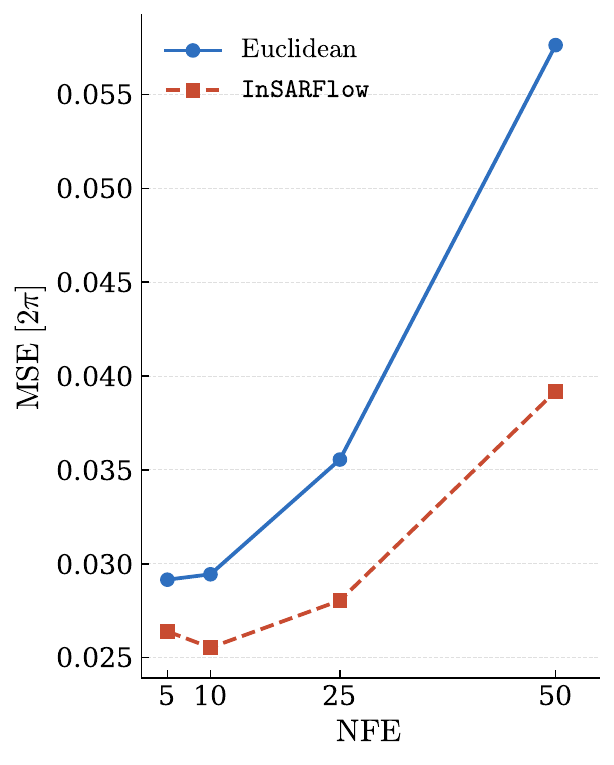}
        \label{fig:mse_nsteps}
    \end{subfigure}
    \hfill
    \begin{subfigure}[b]{0.49\linewidth}
        \centering
        \includegraphics[width=0.8\linewidth]{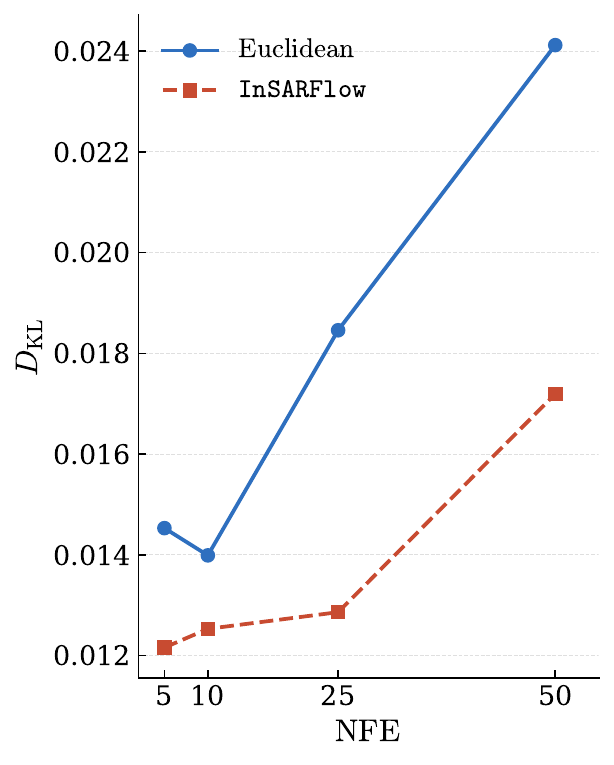}
        \label{fig:kl_nsteps}
    \end{subfigure}
    \caption{MSE $[2\pi]$ (left) and $D_\text{KL}$ (right) as functions of the number of function evaluations NFE for the Euclidean flow matching and \insarflow{} models. \insarflow{} outperforms Euclidian flow matching in all NFE regimes regarding both metrics}
    \label{fig:nsteps}
\end{figure}

\begin{figure*}[t]
\centering
\includegraphics[width=0.7\linewidth]{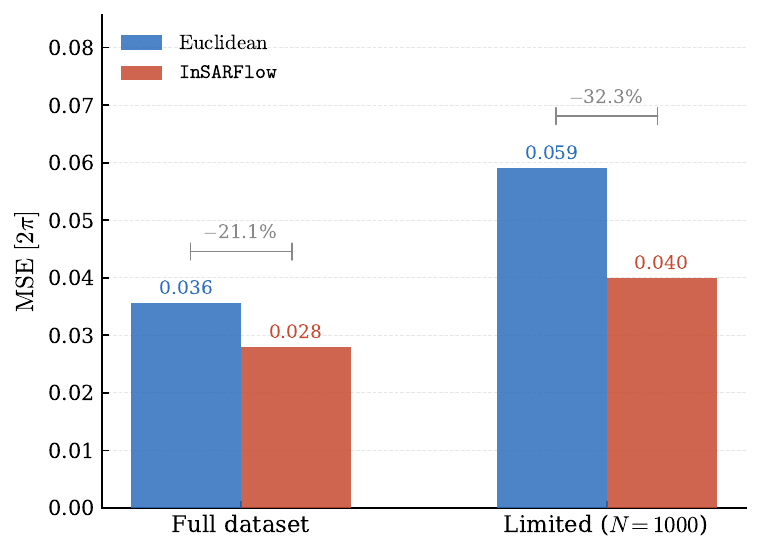}
\caption{Comparison of $\mathrm{MSE}\ [2\pi]$ for the Euclidean and \insarflow{} models on the full dataset and a limited subset ($N=1000$ samples), evaluated at $\mathrm{NFE}=25$. The performance gap widens significantly under data scarcity ($-32.3\%$ vs.\ $-21.1\%$), suggesting that the Riemannian geometric prior becomes increasingly beneficial as training data decreases.}
\label{fig:limited_samples_ablation}
\end{figure*}
\begin{table}[h]
\centering
\caption{%
  \textbf{Quantitative results.}
  \textit{(Left)} Comparison on the Mexico dataset; \insarflow{} outperforms all baselines on both metrics.
  \textit{(Right)} Cross-domain evaluation of two \insarflow{} variants trained on Mexico and simulated datasets (NFE=10).
}
\label{tab:combined}
\setlength{\tabcolsep}{4pt}
\subcaptionbox{Quantitative results on Mexico dataset.\label{tab:mexico_metrics}}{%
{
\begin{tabular}{@{}lcc@{}}
\cmidrule(lr){2-3}
 & \multicolumn{2}{c}{\textbf{Mexico dataset}}\\
\midrule
\textbf{Method} & MSE $[2\pi]\downarrow$ & $D_\text{KL}\downarrow$\\
\midrule
TV denoiser & 0.807 & 0.016\\
UNet        & 0.594 & 0.491\\
\rowcolor{customblue!50} \insarflow{} & \textbf{0.487} & \textbf{0.005}\\
\bottomrule
\end{tabular}%
}
}
\hfill
\subcaptionbox{Cross-domain evaluation (NFE=10).\label{tab:cross_domain}}{%
\begin{tabular}{@{}lcc|cc@{}}
 & \multicolumn{4}{c}{\textbf{Test Domain}}\\
\cmidrule(lr){2-5}
 & \multicolumn{2}{c}{Mexico} & \multicolumn{2}{c}{Simulation}\\
\midrule
\textbf{Model} & MSE$\downarrow$ & $D_\text{KL}\downarrow$ & MSE$\downarrow$ & $D_\text{KL}\downarrow$\\
\midrule
Mexico     & 0.487 & 0.005 & 0.131 & 0.028\\
Simulation & 0.493 & 0.009 & 0.026 & 0.013\\
\bottomrule
\end{tabular}%
}
\end{table}


    In the previous section, we showed the performances of \insarflow{} to denoise images from a simulated dataset. Now, we assess its performances when trained on real-world data, namely on the Mexico City dataset.

The denoising obtained for two different dates (\cref{fig:main_denoised}) is accurate, and \insarflow{} manages to reconstruct the fringes even when the patterns are fine, including in the case where the noisy interferogram is computed between the images with the largest temporal separation ($\Delta t = 468$ days). 
In addition, \insarflow{} predicts denoised interferograms that are visually similar to those produced by the \texttt{COFI-PL} algorithm (see \cref{tab:mexico_metrics}).  
However, we can note that for the challenging case of a large temporal separation case, some complex fringe patterns are not completely retrieved by the method.
On the other hand, we did not notice hallucinations from the model, \ie ~the unexpected apparition of non-existing fringes in the ground truth, which is a desirable feature in this application context.

\subsection{Cross-domain evaluation}

\Cref{tab:cross_domain} reports the cross-domain denoising performance of two \insarflow{} models trained on different data distributions. Each model performs best when evaluated on its corresponding test domain, with the Mexico-trained model achieving a lower cyclical MSE on the Mexico dataset, and the simulation-trained model yielding the best results on the simulated data.
Discrepancies in interferogram statistical properties are responsible for the observed performance degradation in cross-domain evaluations.

\subsection{Ablations}

A major concern with denoising models based on time-dependent processes is their inference time. Indeed, these models typically require multiple function evaluations (NFEs) to sample or denoise new images, with early diffusion models requiring up to 1,000 NFEs. We conduct an ablation study by evaluating denoising performance on the synthetic dataset as a function of the number of inference steps, as shown in \cref{fig:mse_nsteps,fig:kl_nsteps}. We observe that \insarflow{} achieves the best performance at 10 NFEs, increasing the NFE beyond this point leads to error accumulation and results in a higher MSE.

This ablation study thus highlights that only a small number of sampling steps is sufficient to obtain strong denoising performance with \insarflow{}, enabling fast and efficient inference.
Figure~\ref{fig:limited_samples_ablation} shows that \insarflow{} consistently outperforms the Euclidean baseline in terms of circular MSE, with a relative improvement of $-21.1\%$ on the full dataset. Notably, this gap widens to $-32.3\%$ when training data is limited to $N=1000$ samples, suggesting that the Riemannian geometric prior acts as an effective regularizer under data scarcity.



\section{Conclusion}

We introduced \insarflow{}: a method that leverages Riemannian flow matching for either generating or denoising interferograms in InSAR.
This approach inherently integrates a strong prior knowledge about the data structure by treating interferograms as points on the torus $\mathbb{T}_d$.
This notably guarantees that the generated interferograms respect the inherent periodic structure of the phase.
In the generative setting, we showed that this method can generate realistic Sentinel-1 type interferograms, notably by rendering the expected texture and fringe patterns. 
%
%

In the denoising setting, we adopted a distillation approach by transporting the distribution of noisy interferograms to the one of data processed by a multitemporal InSAR method (interferometric phase linking).
This resulted in a method that achieves results comparable to this baseline, even when provided with only a single interferogram.
\insarflow{} has the potential to accelerate the availability of denoised interferograms by reducing the reliance on a complete interferogram time-series.

\section*{Acknowledgement}

This work was conducted as part of the research projects \href{https://mage.science/}{\textsc{MAGE}} (ANR-22-CE23-0010) funded by the \emph{Agence
Nationale de la Recherche}. This work was granted access to the HPC resources of IDRIS under the allocations AD011014327R2 made by GENCI.

\bibliographystyle{iclr2026_conference}
\bibliography{references}

\end{document}